\documentclass[conference, 10pt]{IEEEtran}
\IEEEoverridecommandlockouts
\usepackage[dvips]{graphicx}
\usepackage{color}
\usepackage{times}
\usepackage{fancyhdr}
\usepackage{amssymb,amsmath}
\usepackage{lastpage}
\usepackage{epsfig}
\usepackage{algorithm2e}

\newtheorem{theorem}{Theorem}[section]
\newtheorem{lemma}[theorem]{Lemma}

\def\proof{\noindent\hspace{2em}{\itshape Proof: }}
\def\endproof{\hspace*{\fill}~\QED\par\endtrivlist\unskip}
\def\QEDclosed{\mbox{\rule[0pt]{1.3ex}{1.3ex}}} 

\def\QED{\QEDclosed} 

\makeatletter
\def\ps@headings{%
\def\@oddhead{\mbox{}\scriptsize\rightmark \hfil \thepage}%
\def\@evenhead{\scriptsize\thepage \hfil \leftmark\mbox{}}%
\def\@oddfoot{}%
\def\@evenfoot{}}
\makeatother
\begin{document}
\title{Network Coding for Multi-Resolution Multicast}
\author{\authorblockN{MinJi Kim, Daniel Lucani, Xiaomeng Shi, Fang Zhao, Muriel M\'edard}
\authorblockA{Massachusetts Institute of Technology, Cambridge, MA 02139, USA\\
Email: \{minjikim, dlucani, xshi, zhaof, medard\}@mit.edu}
}
\maketitle

\begin{abstract}
Multi-resolution codes enable multicast at different rates to different receivers, a setup that is often desirable for graphics or video streaming. We propose a simple, distributed, two-stage message passing algorithm to generate network codes for single-source multicast of multi-resolution codes. The goal of this \emph{pushback algorithm} is to maximize the total rate achieved by all receivers, while guaranteeing decodability of the base layer at each receiver. By conducting pushback and code generation stages, this algorithm takes advantage of inter-layer as well as intra-layer coding. Numerical simulations show that in terms of total rate achieved, the pushback algorithm outperforms routing and intra-layer coding schemes, even with codeword sizes as small as 10 bits. In addition, the performance gap widens as the number of receivers and the number of nodes in the network increases. We also observe that naiive inter-layer coding schemes may perform worse than intra-layer schemes under certain network conditions.
\end{abstract}

\section{Introduction}\label{sect:Introduction}
Many real-time applications, such as teleconferencing, video
streaming, and distance learning, require multicast from a single
source to multiple receivers. In conventional multicasts, all
receivers receive at the same rate. In practice, however, receivers
can have widely different characteristics. It becomes desirable to
service each receiver at a rate commensurate with its own demand and
capability. One approach to multirate multicast is to use
multi-description codes (MDC), dividing source data into equally
important streams such that the decoding quality using any subset of
the streams is acceptable, and better quality is obtained by more
descriptions. A popular way to perform MDC is to combine layered
coding with the unequal error protection of a priority encoding
transmission (PET) system \cite{ABEL96}. Another approach for
multirate multicast is to use multi-resolution codes (MRC), encoding
data into a base layer and one or more refinement layers \cite{S92,
effros2001}. Receivers subscribe to the layers cumulatively, with
the layers incrementally combined at the receivers to provide
progressive refinement. The decoding of a higher layer always
requires the correct reception of all lower layers including the
base layer.

In this paper, we consider multirate multicast with linear network
coding. Proposed in \cite{ACLY00}, network coding allows and
encourages mixing of data at intermediate nodes. It has been shown
that in single rate multicast, network coding achieves the minimum
of the maximum flow to each receiver, although this limit is
generally not achievable through traditional routing schemes.
K\"{o}tter and M\'{e}dard also studied multirate multicast, deriving
necessary algebraic conditions for the existence of network coding
solutions for a given network and receiver requests
\cite{koettermedard}. For $n$-layer multicast, linear network codes
can satisfy requests from all the receivers if the $n$ layers are to
be multicasted to all but one receiver. If more than one subscribe
to subsets of the layers, linear codes cease to be sufficient.

Previous work on multirate multicast with network coding includes
\cite{SK07, WS08, wu2008rnf, sundaram2005, zyzz06,  XXZW07}. For the
MDC approach, references \cite{SK07} and \cite{WS08} modified PET at
the source to cater for network coded systems. Recovery of some
layers is guaranteed before full rank linear combinations of all the
layers are received, and this is achieved at the cost of a lower
code rate. Wu \emph{et al.} studied the problem of Rainbow Network
Coding, which incorporates linear network coding into
multi-description coded multicast \cite{wu2008rnf}. For the MRC
approach, Sundaram \emph{et al.} studied multi-resolution media
streaming, and proposed a polynomial-time algorithm for multicast to
heterogeneous receivers \cite{sundaram2005}. Zhao \emph{et al.}
considered multirate multicast in overlay networks \cite{zyzz06}.
They organized receivers into layered data distribution meshes, and
utilized network coding in each mesh.  Xu \emph{et al.} proposed the
Layered Separated Network Coding Scheme to maximize the total number
of layers received by all receivers \cite{XXZW07}.

In the work mentioned above, if no additional coding at the source
such as modified PET is used, the aggregate rate to all receivers is
maximized by solving the linear network coding problem separately
for each layer \cite{ wu2008rnf,sundaram2005, zyzz06, XXZW07}.
Specifically, for each layer, a subgraph is selected for network
coding by performing linear programming. In other words, only
\emph{intra-layer} network coding is allowed. On the other hand,
\emph{inter-layer} network coding, which allows coding across
layers, often achieves higher throughput, and is more powerful.
Incorporation of inter-layer linear network coding into multirate
multicast, however, is significantly more difficult, as intermediate
nodes have to know the network topology and the demands of all
down-stream receivers before determining its network codes.
Reference \cite{DSW09} considers inter-layer network coding by
partitioning  the layers into groups at the source, and performing
``intra-group" coding. If we define these ``groups" as the new
layers, this approach can also be categorized as intra-layer network
coding. On the other hand, the algorithm we propose in this paper
does not impose such grouping, and coding can happen at any node
across any layers. 

In this paper, we propose a simple, distributed, two-stage message
passing algorithm to generate network codes for single source
multicast of multi-resolution codes. Unlike previous work, this
algorithm allows both intra-layer and inter-layer network coding at
all nodes. It guarantees decodability of the base layer at all
receivers. In terms of total rate achieved, it outperforms routing
as well as network coding schemes that involve intra but not
inter-layer coding, with field size as small as $2^{10}$. The
performance gain of this algorithm increases as the number of
receivers increases and as the network grows in size, if appropriate
criterion is used. Otherwise,  na\"ive inter-layer coding may lead to an inappropriate choice of network code, which can be
worse than intra-layer network coding.

The rest of this paper is organized as follows. A network model and
the network coding problem of multicast of multi-resolution codes
are established in Section \ref{sec:problemsetup}. The pushback
algorithm is proposed in Section \ref{sec:pushback}, and proved in
Section \ref{sec:analysis} to guarantee decodability of the base
layer. Simulation results are presented in Section
\ref{sec:simulations}, while discussions on future work conclude the
paper in Section \ref{sec:conclusions}.

\section{Problem Setup}\label{sec:problemsetup}
\begin{figure}[tbp]
\begin{center}
\includegraphics[width=0.40\textwidth]{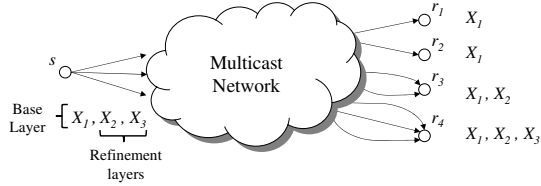}
\end{center}\vspace*{-.5cm}
\caption{A network with a source $s$ with multi-resolution codes
$\mathcal{X}_1, \mathcal{X}_2$, and $\mathcal{X}_3$, and receivers
$r_1, r_2, r_3, r_4$.}\vspace*{-.3cm}\label{fig:network}
\end{figure}

We consider the network coding problem for single-source multicast
of multi-resolution codes, as illustrated by
Figure~\ref{fig:network}. The single-source multicast network of
interest is modeled by a directed acyclic graph $G = (V, E)$, $V$
being the set of nodes, and $E$ the set of links. Each link is
assumed to have unit capacity, while links with capacities greater
than 1 are modeled with multiple parallel links. The subset $R =
\{r_1, r_2, ... r_n\} \subseteq V$ is the set of receivers which
wish to receive information from the source node $s \in V$. The
source processes, $\mathcal{X}_1, \mathcal{X}_2, ...,
\mathcal{X}_L$, constitute a multi-resolution code, where
$\mathcal{X}_1$ is the \emph{base layer} and the rest are the
\emph{refinement layers}. It is important to note that layer
$\mathcal{X}_i$ without layers $\mathcal{X}_1,
\mathcal{X}_2,...,\mathcal{X}_{i-1}$ is not useful for any $i$. For
simplicity, we assume each layer is of unit rate. Therefore, given a
link $e \in E$, we can transmit one layer (or equivalent coded data
rate) on $e$ at a time. The min-cut between $s$ and a node $v$ is
denoted by $minCut(v)$,  and we assume that every node $v$ knows its
$minCut(v)$. Note that there are efficient algorithms, such as
Ford-Fulkerson algorithm, that can compute $minCut(v)$.

Our goal is to design a simple and distributed algorithm that
provides a coding strategy to maximize the total rate achieved by
all receivers with the reception of the base layer guaranteed to all
receivers. By Min-Cut Max-Flow bound, each receiver $r_i$ can
receive at most $minCut(r_i)$ layers ($\mathcal{X}_1$,
$\mathcal{X}_2$, .., $\mathcal{X}_{minCut(r_i)}$). We present the
\emph{pushback} algorithm, and compare its performance against other
existing algorithms and against the theoretical bound of Min-Cut
Max-Flow.

\section{Pushback Algorithm}\label{sec:pushback}
\begin{figure}[tbp]
\begin{center}\vspace*{-0.5cm}
\includegraphics[width=0.47\textwidth]{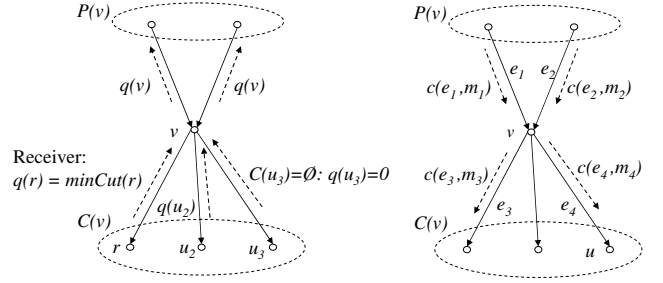}
\end{center}\vspace*{-0.5cm} \caption{Pushback stage and code assignment stage
at node $v$.}\vspace{-0.3cm}\label{fig:pushback}
\end{figure}

The pushback algorithm is a distributed algorithm which allows both
intra-layer and inter-layer linear network coding. It consists of
two stages: \emph{pushback} and \emph{code assignment}.

In the pushback stage, messages initiated by the receivers are
pushed up to the source, allowing upstream nodes to gather
information on the demand of any receiver reachable from them.
Messages are passed from nodes to their parents. Initially, each
receiver $r_i \in R$ requests for layers $\mathcal{X}_1,
\mathcal{X}_2, ..., \mathcal{X}_{minCut(r_i)}$ to its upstream
nodes, \emph{i.e.}, the receiver $r_i$ requests to receive at a rate
equal to its min-cut. An intermediate node $v \in V$ computes a
message, which depends on the value of $minCut(v)$ and the requests
from its children. Node $v$ then pushes this message to its parents,
indicating the layers which the parent node should
encode together. 

The code assignment stage is initiated by the source once pushback stage
is completed. Random linear network codes \cite{HKME06} are
generated in a top-down fashion according to the pushback messages.
The source $s$ generates codes according to the messages from its children: $s$ encodes the requested layers together and
transmits the encoded data to the corresponding child. Intermediate
nodes then encode/decode the packets according to the messages
determined during pushback.

To describe the algorithm formally, we introduce some additional
notations. For a node $v$, let $P(v)$ be its set of parent nodes,
and $C(v)$ its children as shown in Figure~\ref{fig:pushback}.
$P(v)$ and $C(v)$ are disjoint since the graph is acyclic. Let
$E^{in}_v = \{(v_1, v_2) \in E\ |\ v_2 = v\}$ be the set of incoming
links, and $E^{out}_v = \{(v_1, v_2) \in E\ |\ v_1 = v\}$ the set of
outgoing links.

\subsection{Pushback Stage}
\begin{algorithm}[bp]
\vspace*{-0.5cm} \SetLine \If{$v$ is a receiver}{ $q(v) =
minCut(v)$\; } \If{$v$ is an intermediate node}{ \If{$C(v) =
\emptyset$}{ $q(v) = 0$\; } \If{$C(v) \ne \emptyset$}{ $q(v) =
f(q(C(v)), minCut(v))$\; } } \caption{The pushback stage at node
$v$.}\label{alg:pushback}
\end{algorithm}
As shown in Figure~\ref{fig:pushback}, we denote  the message
received by node $v$ from a child $u \in C(v)$ as $q(u)$, and the
set of messages received by node $v$ from its children as $q(C(v)) =
\{q(u) | u\in C(v) \}$. A message $q(u)$ means that $u$ requests its
parents to code across layers 1 to \emph{at most} $q(u)$. Once
requests are received from all of its children, $v$ computes its message $q(v)$ and sends the same
message $q(v)$ to all of its parents. The request $q(v)$ is a
function of $q(C(v))$ and $minCut(v)$, \emph{i.e.} $q(v) =
f(q(C(v)), minCut(v))$. A pseudocode for the pushback stage at a
node $v\in V$ is shown in Algorithm \ref{alg:pushback}. It is
important to note the choice of $f(\cdot)$ is a key feature of the
algorithm as it determines the performance. We present two different
versions of $f(\cdot)$: \emph{min-req criterion} and \emph{min-cut
criterion}, which we discuss next.

\subsubsection{Min-req Criterion}

The min-req criterion, as the name suggests, defines $f(\cdot)$ as follows:
\begin{align*}
q(v) &= f(q(C(v)), minCut(v))\\
& = \begin{cases}
0 & \text{if $q(u) = 0$ for all $u \in C(v)$,}\\
q_{min} &  \text{otherwise,}
\end{cases}
\end{align*}
where $q_{min} = \min_{q(u) \ne 0,\ u \in C(v)} q(u)$ is the minimum
non-zero $q(u)$ from $u \in C(v)$.

\begin{figure}[tbp]
\begin{center}
\includegraphics[width=0.45\textwidth]{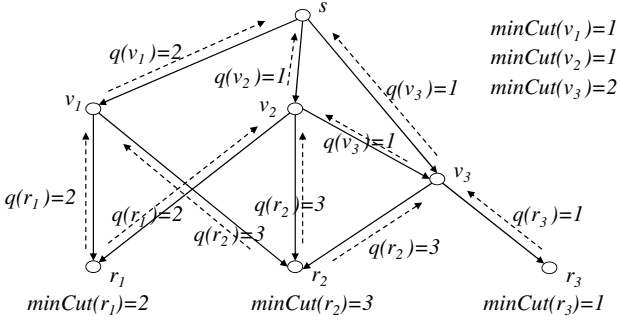}
\end{center}\vspace*{-.5cm}
\caption{An example of pushback stage with min-req
criterion.}\label{fig:pushbackMinReq}
\end{figure}

This criterion may seem very pessimistic and na\"ive, as the
intermediate nodes serve only the minimum requested by their
downstream receivers to ensure the decodability of the base layer.
Nonetheless, as we shall see in Section \ref{sec:simulations}, the
performance of this criterion is quite good. An example of pushback
with min-req is shown in Figure \ref{fig:pushbackMinReq}. Receivers
$r_1$, $r_2$, and $r_3$ request their min-cut values 2, 3, and 1,
respectively. The intermediate nodes $v_1$, $v_2$, and $v_3$ request
the minimum of all the requests received, which are 2, 1, and 1,
respectively.
\begin{figure}[tbp]
\begin{center}
\vspace*{.4cm}
\includegraphics[width=0.45\textwidth]{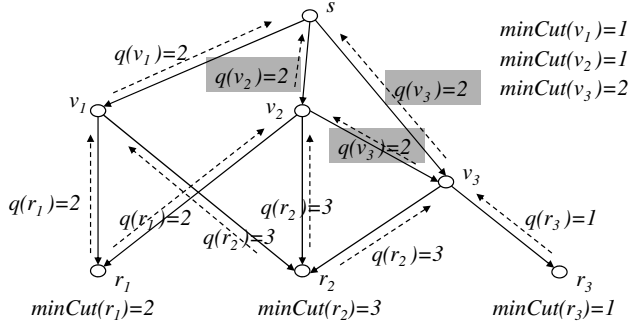}
\end{center}\vspace*{-.5cm}
\caption{An example of pushback stage with min-cut criterion.
Highlighted are the messages different from that of min-req criterion
(Figure
\ref{fig:pushbackMinReq}).}\vspace*{-.3cm}\label{fig:pushbackMinCut}
\end{figure}

\subsubsection{Min-cut Criterion}
The min-cut criterion defines the function $f(\cdot)$ as follows:
\begin{align*}
q(v) &= f(q(C(v)), minCut(v))\\
&=
\begin{cases}
q_{min} & \text{if $minCut(v) \leq q_{min}$,}\\
minCut(v) & \text{otherwise,}
\end{cases}
\end{align*}
where $q_{min} = \min_{q(u) \ne 0,\ u \in C(v)} q(u)$.

Note if a node $v$ receives $minCut(v)$ number of linearly
independent packets coded across layers 1 to $minCut(v)$, it can
decode layers $\mathcal{X}_1, \mathcal{X}_2, ...,
\mathcal{X}_{minCut(v)}$ and act as a secondary source with those
layers. Thus, if there is at least one child $u \in C(v)$ that
requests fewer than $minCut(v)$ layers, \emph{i.e.}  $minCut(v) >
q_{min}$, node $v$ sets its request $q(v)$ to $minCut(v)$. However,
if all nodes request more than $minCut(v)$ layers, node $v$ does not
have sufficient capacity to decode the layers requested by its
children. Thus, it sets $q(v)=q_{min}$. An example of pushback with
min-cut is shown in Figure~\ref{fig:pushbackMinCut}. The network is
identical to that of Figure~\ref{fig:pushbackMinReq}. Again, the
nodes $r_1$, $r_2$, $r_3$, and $v_1$ request 2, 3, 1, and 2,
respectively. However, node $v_2$ requests 2, which is the minimum
of all the requests it received, and node $v_3$ requests
$minCut(v_3) = 2$.

\subsection{Code Assignment Stage}
\begin{algorithm}[tbp]
\SetLine
\If{$v$ is the source $s$}{
\ForEach{edge $e = (v, u) \in E^{out}_v$}{
$v$ transmits $c(e, q(u))$\;
}
}
\If{$v$ is an intermediate node}{
\If{$P(v) = \emptyset$}{
$v$ sets $c(e, 0)$ for all $e \in E^{out}_v$\;
}
\If{$P(v) \ne \emptyset$}{
$v$ receives codes $c(e_i, m_i)$, $e_i \in E^{in}_v$\;
$v$ determines $m^*$, which is the maximum $m$ such that $\mathcal{X}_1, \mathcal{X}_2, ... \mathcal{X}_{m}$ are decodable from $c(e_i, m_i)$'s\;
\ForEach{child $u \in C(v)$}{
Let $e = (v,u)$\;
\If{$q(u) \leq m^*$}{
$v$ decodes layers $\mathcal{X}_1, \mathcal{X}_2, ..., \mathcal{X}_{m^*}$\;
$v$ transmits $c(e, q(u))$\;
}
\If{$q(u) > m^*$}{
Let $m_{max} = \max_{m_i \leq q(u)} m_i$\;
$v$ transmits $c(e, m_{max})$\;
}}}}
\If{$v$ is a receiver}{
$v$ receives codes $c(e_i, m_i)$, $e_i \in E^{in}_v$\;
$v$ decodes $m^*$ layers, which is the maximum $m$ such that $\mathcal{X}_1, \mathcal{X}_2, ... \mathcal{X}_{m}$ are decodable from $c(e_i, m_i)$'s\;
}\caption{The code assignment stage at node $v$.}\label{alg:codeassignment}
\end{algorithm}
This stage is initiated by the source after pushback is completed.
As shown in Figure~\ref{fig:pushback}, $c(e,m)$ denotes the random
linear network code $v$ transmits to its child $u \in C(v)$, where
$e = (v, u)$, and $m$ means that packets on $e$ are coded across
layers 1 to $m$. Note $m$ may not equal to $q(u)$, which we discuss
in more detail in Section \ref{sec:analysis}. Algorithm
\ref{alg:codeassignment} presents a pseudocode for the code
assignment stage at any node $v\in V$.

Algorithm~\ref{alg:codeassignment} considers source, intermediate,
and receiver nodes separately. The source always exactly satisfies
any requests from its children, while the receivers decode as many
consecutive layers as they can. For an intermediate node $v$
connected to the network $(P(v) \ne \emptyset)$, $v$ collects all
the codes $c(e_i, m_i)$ from its parents and determines $m^*$, the
number of layers up to which $v$ can decode. It is possible that $v$
cannot decode any layer, leading to an $m^*$ equal to zero. For $m^*
\ne 0$, $v$ can act as a secondary source for layers 1, 2, ...,
$m^*$ by decoding these layers. In the case where $q(u) \leq m^*$, $u \in C(v)$, $v$ can
satisfy $u$'s request exactly by encoding just the layers 1 to
$q(u)$. If $q(u) > m^*$, $v$ cannot decode the layers $u$ requested;
thus, cannot satisfy $u$'s request exactly. Therefore, $v$ uses a
best effort approach and delivers a packet coded across $m_{max}$
layers, where $m_{max}$ is the closest to $q(u)$ node $v$ can serve without violating
$u$'s request, \emph{i.e.} $q(u) \geq m_{max}$. The code assignment stage requires that every node
checks its decodability to determine $m^*$. This process involves
Gauss-Jordan elimination, which is computationally cheaper than
matrix inversion required for decoding. Note that only a subset of
the nodes need to perform (partial) decoding.

\begin{figure}[tbp]
\begin{center}
\includegraphics[width=0.45\textwidth]{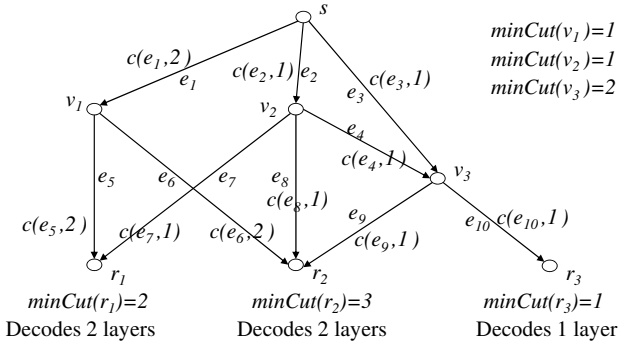}
\end{center}\vspace*{-.5cm}
\caption{An example of code assignment stage with min-req
criterion.}\label{fig:codeMinReq}
\end{figure}

\begin{figure}[tbp]
\begin{center}
\includegraphics[width=0.45\textwidth]{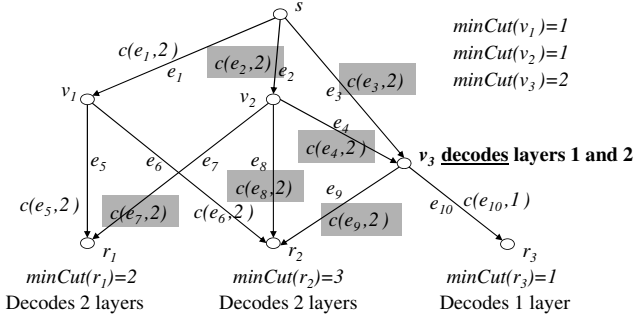}
\end{center}\vspace*{-.5cm}
\caption{An example of code assignment stage with min-cut criterion.
High--lighted are the messages different from the min-req
criterion (Figure \ref{fig:codeMinReq}).}\label{fig:codeMinCut}
\end{figure}

Figures~\ref{fig:codeMinReq} and \ref{fig:codeMinCut} illustrate the
code assignment stage for the examples in Figures
\ref{fig:pushbackMinReq} and \ref{fig:pushbackMinCut}, respectively.
Note that the algorithm for code assignment stays the same, whether
we use min-req or min-cut  criterion during the pushback stage;
however, the resulting code assignments are different. Although the
throughput achieved in this example network in  Figures
\ref{fig:codeMinReq} and \ref{fig:codeMinCut} are the same, this is
usually not the case. Generally the min-cut criterion achieves
higher throughput than the  min-req criterion.

\section{Analysis of Pushback Algorithm}\label{sec:analysis}
In general, not all receivers can achieve their min-cuts through
linear network coding. Nonetheless, we want to guarantee that no
receiver is denied service, \emph{i.e.} although some nodes may not
receive up to the number of layers they requested, all should
receive at least layer 1. In this section, we prove that the
pushback algorithm guarantees decodability of the base layer,
$\mathcal{X}_1$, at all receivers. Two related
lemmas 
are presented to prove Theorem \ref{thm:main}.

\begin{lemma}\label{thm:baselayer}
Assume $minCut(v) = n$ for a node $v$ in $G$. In the pushback
algorithm, if $m_i \leq n$ for all $c(e_i, m_i)$, $e_i \in
E^{in}_v$, then $v$ can decode at least layer 1 with high
probability. In other words, if all received codes at $v$ are
combinations of at most $n$ layers, $v$ can decode at least layer 1.
\end{lemma}
\proof Recall that in the
pushback algorithm, a code $c(e_i,m_i)$ represents coding across
layers 1 to $m_i$; if the field size is large, with high
probability, the first $m_i$ elements of this coding vector are
non-zero, whereas the rest are zeros.

\begin{figure}[tbp]
\begin{center}\vspace*{-.5cm}
\includegraphics[width=0.42\textwidth]{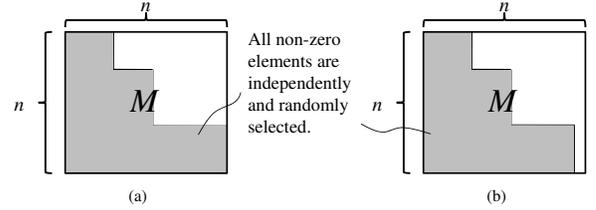}
\end{center}\vspace*{-.5cm}
\caption{Coding matrix $M$; each row represents a code received, and
columns represent the layers. The maximum number of non-zero columns
in $M$, $c_k$, can be equal to $n$ (as shown in (a)), or less than
$n$ (as shown in (b)).}\vspace*{-.5cm}\label{fig:matrix}
\end{figure}

Since $minCut(v) = n$, there exist $n$
edge-disjoint paths from the source $s$ to $v$, for all links are
assumed to have unit capacity. Therefore, $v$ receives from its incoming links
at least $n$ codes, each of which can be represented as a row coding vector of length $n$, since $m_i < n$ for all $i$.
We pick the $n$ codes corresponding to the edge-disjoint
paths to obtain an $n\times n$ coding matrix. For the square coding matrix,
we sort its rows according to the number of non-zero elements per
row, obtaining the structure shown in Figure \ref{fig:matrix}. We
denote this sorted matrix by $M$, and the unique numbers of non-zero
elements in its rows by $c_1, c_2, ..., c_k$, in ascending order.
Since the rows of $M$ are generated along edge-disjoint paths
using random linear network coding, the non-zero elements of $M$ are
independently and randomly selected.

\begin{figure}[b]
\begin{center}\vspace*{-.5cm}
\includegraphics[width=0.49\textwidth]{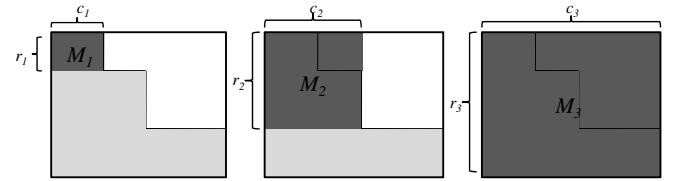}
\end{center}\vspace*{-.5cm}
\caption{Upper-left corner submatrices $M_1$, $M_2$, and
$M_3$.}\label{fig:submatrix}
\end{figure}
Next, we define upper-left corner submatrices $M_1$, $M_2$, ...,
$M_k$ as shown in Figure \ref{fig:submatrix}, where each submatrix
$M_i$ is of size $r_i \times c_i$. More specifically, the rows of
$M$ with exactly $c_1$ non-zero elements form a $r_1 \times
c_1$ submatrix $M_1$; the rows of $M$ with exactly $c_1$ or $c_2$
non-zero elements form the $r_2 \times c_2$ submatrix $M_2$. $M_k$
is of size $r_k\times c_k$, where $r_k=n$, and $c_k\le n$. Note for
any $M_i$ of those submatrices, if $rank(M_i) = c_i$, node $v$ can
decode layers 1 to $c_i$, \emph{i.e.}, the base layer is decodable.
In other words, layer 1 is not decodable at node $v$ only if
$rank(M_i) < c_i$ for all $i$.

With these definitions, we assume layer 1 is not decodable at node
$v$, and prove the lemma by contradiction. Specifically, we prove by
induction that this assumption implies $r_i <  c_i$ for all $i$,
leading to the contradiction $r_k < c_k$.

For the base case, first consider $M_1$. If layer 1 is not
decodable, $rank(M_1) < c_1$. Recall that elements in $M_1$ are
independently and randomly selected \cite{HKME06}; if $r_1 \geq
c_1$, with high probability, $rank(M_1) = c_1$. Therefore, the above
assumption implies $r_1 < c_1$ and $rank(M_1) = r_1$. Next consider
$M_2$. Under the assumption that layer 1 is not decodable,
$rank(M_2) < c_2$. Since $rank(M_1) = r_1$ and $M_2$ includes rows
of $M_1$, $rank(M_2) \geq r_1$. Rows $r_1 + 1, r_1 + 2, ..., r_2$
are called the \emph{additional rows} introduced in $M_2$. If there are more than $c_2 -r_1$ additional rows, $M_2$ has full
rank, \emph{i.e.} $rank(M_2) = c_2$, with high probability. Hence,
the number of additional rows in $M_2$ must be less than $c_2-r_1$,
implying $r_2 < c_2$.

For the inductive step, consider $M_i$, $3\leq i\leq k$. Assume that
$r_j < c_j$ for all $j < i$. If layer 1 is not decodable, $rank(M_i)
< c_i$. By similar arguments as above, $rank(M_{i-1}) = r_{i-1}$,
and there must be less than $c_i - r_{i-1}$ additional rows
introduced in $M_i$. Thus, $r_i < c_i$. By induction, the total
number of rows $r_k = n$ in $M$ is strictly less than $c_k\leq n$, which is
a contradiction. We therefore conclude that node $v$ can decode the
base layer. In fact, $v$ can decode at least $c_1$ layers.
\endproof

\begin{figure}[tbp]
\begin{center}
\includegraphics[width=0.16\textwidth]{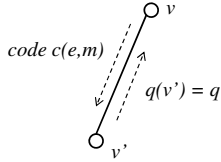}
\end{center}\vspace*{-.5cm}
\caption{Node $v$ and its child
$v'$.}\vspace*{-.3cm}\label{fig:request}
\end{figure}

\begin{figure}[bp]
\begin{center}\vspace*{-.5cm}
\includegraphics[width=0.35\textwidth]{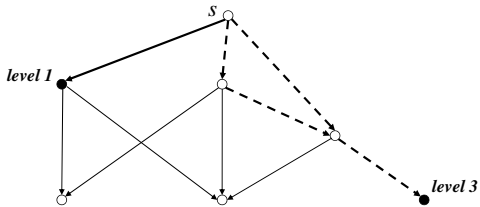}
\end{center}\vspace*{-.5cm}
\caption{Level 1 and level 3 node in a network.}\label{fig:level}
\end{figure}

\begin{lemma}\label{thm:mincut}
In the pushback algorithm, for each link $e = (v, v')$, assume that
node $v'$ sends request $q(v') = q$ to node $v$. Then, the code
$c(e, m)$ on link $e$ is coded across at most $q$ layers,
\emph{i.e.} $m \leq q$ (see Figure~\ref{fig:request}).
\end{lemma}
\proof First, we define the notion of \emph{levels}. A node $u$ is in
level $i$ if the longest path from $s$ to $u$ is $i$, as shown in
Figure~\ref{fig:level}. Since the graph is acyclic, each node has a
finite level number. We shall use induction on the levels to prove
that this lemma holds for both min-req and min-cut criteria.

For the base case, if $v'$ is in level 1, it is directly
connected to the source, and receives a code across exactly $q$
layers on $e$ from $s$. For the inductive
step, assume that all nodes in levels 1 to $i$, $1 \leq i < k$, get
packets coded across layers 1 to at most their request. Assume $v'$
is in level $i+1$ as in Figure \ref{fig:request}. Let $v\in P(v')$; therefore, $v$ is in level $j\le i$. Let
$q_{min}$ be the smallest non-zero request at $v$, that is $q_{min}
= min_{q(u) \ne 0, u\in C(v)} q(u)$.

For the min-req criterion, $v$ always sends request $q(v) = q_{min}$
to its parents, and the codes $v$ receives are linear combinations
of at most $q_{min}$ layers. Therefore, the code $v$ sends to its
children is coded across at most $q_{min}$ layers, where $q = q(v')
\geq q_{min}$. In other words, the code received by $v'$ is coded
across at most $q$ layers.

For the min-cut criterion, if $q_{min} > minCut(v)$, node $v$
requests $q(v) = q_{min}$ to its parents. By the same argument as
that for the min-req criterion, $v'$ receives packets coded across
at most $q$ layers. If $q_{min} \leq minCut(v)$, $v$ requests $q(v)
= minCut(v)$. According to the code assignment stage, if $v$ cannot
satisfy request $q$ exactly, it will send out a linear combination
of the layers it can decode. Since $v$ is in level $j \le i$, $v$
receives codes  across layers 1 to at most $minCut(v)$. By Lemma
\ref{thm:baselayer}, node $v$ can decode at least the base layer.
Thus, we conclude that node $v$ is always able to generate a code
for node $v'$ such that it is coded across layers 1 to at most $q$.
\endproof

\begin{theorem}\label{thm:main} In the pushback algorithm, every receiver can decode at least the base layer.
\end{theorem}
\proof The receiver with min-cut $n$ receives linear combination of
at most $n$ layers by Lemma \ref{thm:mincut}. From Lemma
\ref{thm:baselayer}, the receiver, therefore, can decode at least
the base layer.
\endproof

\section{Simulations}\label{sec:simulations}
To evaluate the effectiveness of the pushback algorithm, we
implemented it in Matlab, and compared the performance with both
routing and intra-layered network coding schemes. Random networks
were generated, with a fixed number of receivers  randomly selected
from the vertex set. We consider two metrics to evaluate the
performance:
\begin{align*}
&\small\textit{{\% Happy\ Nodes}} = \frac{\scriptsize{\text{100}}}{\text{\scriptsize{\# of trials}}} \sum_{\text{\tiny{all trials}}} \frac{\text{ \footnotesize{\# of receivers that achieve min-cut}}}{\text{\footnotesize{\# of receivers}}}\,,\\
&\small\textit{{\% Rate\ Achieved}} = \scriptsize{\text{100}} \,\,
\frac{\sum_{\text{\tiny{all trials}}}\text{\footnotesize{total rate
achieved}}}{\sum_{\text{\tiny{all trials}}}
\text{\footnotesize{total min-cut}}}.
\end{align*}
The \textit{\% Happy\ Nodes} metric is the average of the percentage
of receivers that achieved their min-cut, \emph{i.e.} receivers that
received the service they requested. The \textit{\% Rate Achieved}
metric gives a measure of what percentage of the total requested
rate was delivered to the receivers over all trials.

As an example, consider two possible cases where the (min-cut,
achieved-rates) pairs are $([1,1,2],[1,1,1])$ and
$([2,2,3],[2,2,2])$. In both cases, the demand of a single receiver
is missed by one layer, but the corresponding fractions of rates
achieved are $3/4$ and $6/7$ respectively. Using only the \textit{\%
Happy\ Nodes} metric would tell us that $1/3$ of the receivers did
not received all requested layers. However, the \textit{\% Rate
Achieved} metric provides a more accurate measure of how `unhappy' the overall network is.

\subsection{Algorithms for comparison}
\subsubsection{Point-to-point Routing Algorithm}
the point-to-point routing algorithm considers each  multicast as a
set of unicasts. The source node $s$ first multicasts the base layer
$\mathcal{X}_1$ to all receivers. To determine the links used for
layer $\mathcal{X}_1$, $s$ computes the shortest path to each of the
receivers separately. Given the shortest paths to all receivers, $s$
then takes the union of the paths and uses all the links in this
union to transmit the base layer. Note the shortest path to receiver
$r_i$ may not be disjoint with the shortest path to receiver $r_j$.

After transmitting layer $\mathcal{X}_{i-1}$, $2\leq i \leq L$, the
source $s$ uses the remaining network capacity to transmit the next
refinement layer $\mathcal{X}_i$ to as many receivers as possible.
First, $s$ updates the min-cut to all receivers and identifies
receivers that can receive $\mathcal{X}_i$. Let $R' = \{r_{i_1}, r_{i_2},
...\}$ be the set of receivers with updated min-cut greater than 1 and,
therefore, can receive layer $\mathcal{X}_i$. The source $s$ then computes
the shortest paths to receivers in $R'$. The union of these paths is used to transmit the refinement
layer. Node $s$ repeats this process until no receiver can be reached or
there are no more layers to transmit.

\subsubsection{Steiner Tree Routing Algorithm}
the Steiner tree  routing algorithm computes the minimal-cost tree
connecting the source $s$ and all the receivers. We assume that each
link is of unit cost. For the base layer $\mathcal{X}_1$, $s$
computes and transmits on the Steiner tree connecting to all
receivers. For each new refinement layer $\mathcal{X}_i$, s computes
a new Steiner tree to receivers with updated min-cuts greater than
zero. Node $s$ repeats this process to transmit more refinement layers
until no receiver can be reached or the layers are exhausted.

It is important to note that Steiner tree routing algorithm is an
optimal routing algorithm -- it uses the fewest number of links to
transmit each layer. Unlike the point-to-point algorithm, this
algorithm may make routing decisions that is not optimal to any
single receiver, \emph{i.e.} the source may use a non-shortest path
to communicate to a receiver, but it uses fewer links globally.
However, this optimality comes with a cost: the problem of finding a
Steiner tree is NP-complete.

\subsubsection{Intra-layer Network Coding Algorithm}
the intra-layer network coding algorithm uses linear coding on each
layer separately. It iteratively solves the linear programming
problem for linear network coding for layer $\mathcal{X}_i$ with
receivers $R_i = \{r \in R\ |\ minCut(r) \geq 1\}$, where $i = 1$ and $R_1 = R$ initially \cite{lun06}. After
solving the linear program for layer $\mathcal{X}_i$, the algorithm increments $i$,
updates the link capacities, and performs the next round of linear
programming. References \cite{wu2008rnf} and \cite{sundaram2005} are
examples of this intra-layer coding approach.

\subsection{Implementation of Pushback Algorithm}\label{sec:implementation}
The pushback algorithm was implemented with two different message
passing schedules.

\begin{enumerate}
\item Sequential Message Passing: for the pushback stage, each node in the network sends
a request to its parents after request messages from all its children have
been received. For the code assignment stage, each node sends a code to its children after receiving codes from all its parents. This schedule corresponds to the
algorithms explained in Section~\ref{sec:pushback}.
\item Flooding: for the pushback stage, each node updates its request to its parents upon
reception of a new message from its children. For the code assignment stage, each node sends a new code to its children after receiving a new message from any of its parent nodes. This  allows an update
 mechanism that converges to the same solution as Sequential Message
 Passing. In fact, the convergence time depends on the diameter of
 the graph.
\end{enumerate}

Another important issue is the procedure to check decodability at
each node. In general, Gauss-Jordan elimination on the coding matrix in field of size $p$, $\mathbb{F}_p$, is necessary to determine which layers can be decoded at a node after the codes are assigned. However, this is not the case for 2-layer multi-resolution codes.
We define \emph{pattern of coding coefficients} for a node with $L$ incoming links as
$[a_1,a_2,...,a_L]$, where $a_i$ represents the number of layers
combined in the $i$-th incoming link for that node.
If a node receives only the base layer on all incoming links, \emph{i.e.} the pattern of coding coefficients is $[1,1,...,1]$, it can decode this single layer immediately. If at least one of the incoming links contains a
combination of two layers, \emph{i.e.} the pattern of coding
coefficients is one the the following: ${[1,...,1,2], [1,...,1,2,2],
..., [1,2,...,2], [2,...,2]}$, both layers can be decoded as well.
In other words, if there are only two layers, the pattern of coding
coefficients indicates decodability. We note that using the pattern of coding coefficients for decodability is equivalent to using Gauss-Jordan elimination with infinite field size.

In more general cases with more than 2 layers, the pattern of coding
coefficients is not sufficient to determine decodability. For
example, a node with $4$ incoming links of unit rate can have a
min-cut of at most $4$. Assume that a node with $4$ incoming links
has a min-cut of $3$, and that this node is assigned a
coding-coefficient pattern of $[1,1,3,3]$. If all coding vectors are
linearly independent, all layers are decodable.
However, it is possible that the third and the fourth links, both
combining three layers, are not from disjoint paths, \emph{i.e.}
they provide linearly dependent combinations. In such cases, Gauss-Jordan elimination is
necessary to check that only the first layer is decodable.

In subsequent sections, we present simulation results for 2 and
3-layer multi-resolution codes. However, our algorithm is not
limited to 2 and 3-layers; it can be applied to general $n$-layer
multi-resolution codes.

\subsection{Simulation results for 2-layer multi-resolution code}

The simulations for 2-layer multi-resolution code were carried out
for random directed acyclic networks. We averaged 1000 trials for each data
point on the curves plotted in this section. The networks were generated such that the min-cuts and the
in-degrees of all nodes were less than or equal to $2$.

\begin{figure}[tbp]
\begin{center}
\includegraphics[width=0.47\textwidth]{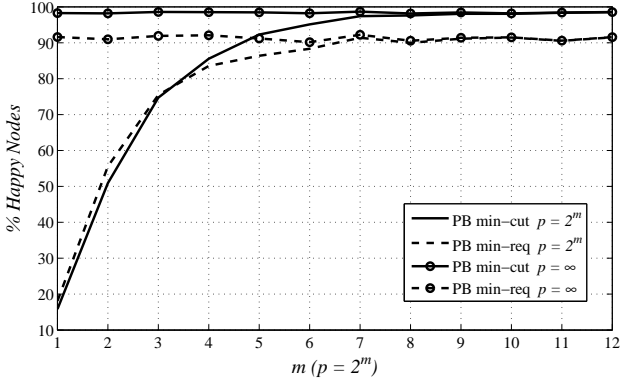}
\end{center}\vspace*{-.4cm}
\caption{Varying field size in a network with 5 receivers and 25
nodes.}\label{fig:layer2nodes25rx5}\vspace*{-.3cm}
\end{figure}

As mentioned in Section \ref{sec:implementation}, the patterns of coding
coefficients are sufficient to check decodability for 2-layer multi-resolution codes, and it is
equivalent to using Gauss-Jordan elimination with an infinite field size. In Figure \ref{fig:layer2nodes25rx5}, we study the effect of field size in a network with 25 nodes and 5 receivers by performing Gauss-Jordan elimination at every node during the code generation
stage with varying field size $p$. Figure~\ref{fig:layer2nodes25rx5} shows the average performance in
terms of \textit{\%~Happy~Nodes} of our pushback algorithm with the
min-cut and min-req criteria against that of using the pattern of coding
coefficients to check decodability. In essence, we are comparing the performance of our system using specific field sizes to that of an infinite field size. It is important to note that even for moderately small field sizes, such as $p \geq 2^8$, the pushback algorithm performs close to that of the system operating at an infinite field size.


Simulation results also illustrate that the min-cut criterion performs considerably better than the min-req criterion for large field sizes, as shown in Figure \ref{fig:layer2nodes25rx5}. However, for small field sizes ($p\leq 2^3$), the min-req criterion is slightly better. This is because the min-req criterion forwards the minimum of the requests received at any node. In the case of a 2-layer multicast, there will be more nodes requesting only the base layer in a network using the min-req criterion than that using the min-cut criterion. Thus, the network using the min-req criterion will have more links carrying only the base layer, which helps improve redundancy for the receivers. This allows several paths to carry the same information, ensuring the decodability of the first layer at the receivers. By comparison, the min-cut criterion tries to combine both layers at as many links as possible. When the field size is large, both layers are decodable with high probability; however, when the field size is small, the probability of generating linearly dependent codes is high. As a result, when $p$ is small, this mixing can prevent decodability of both layers at a subset of receivers.

\begin{figure}[tbp]
\begin{center}
\includegraphics[width=0.47\textwidth]{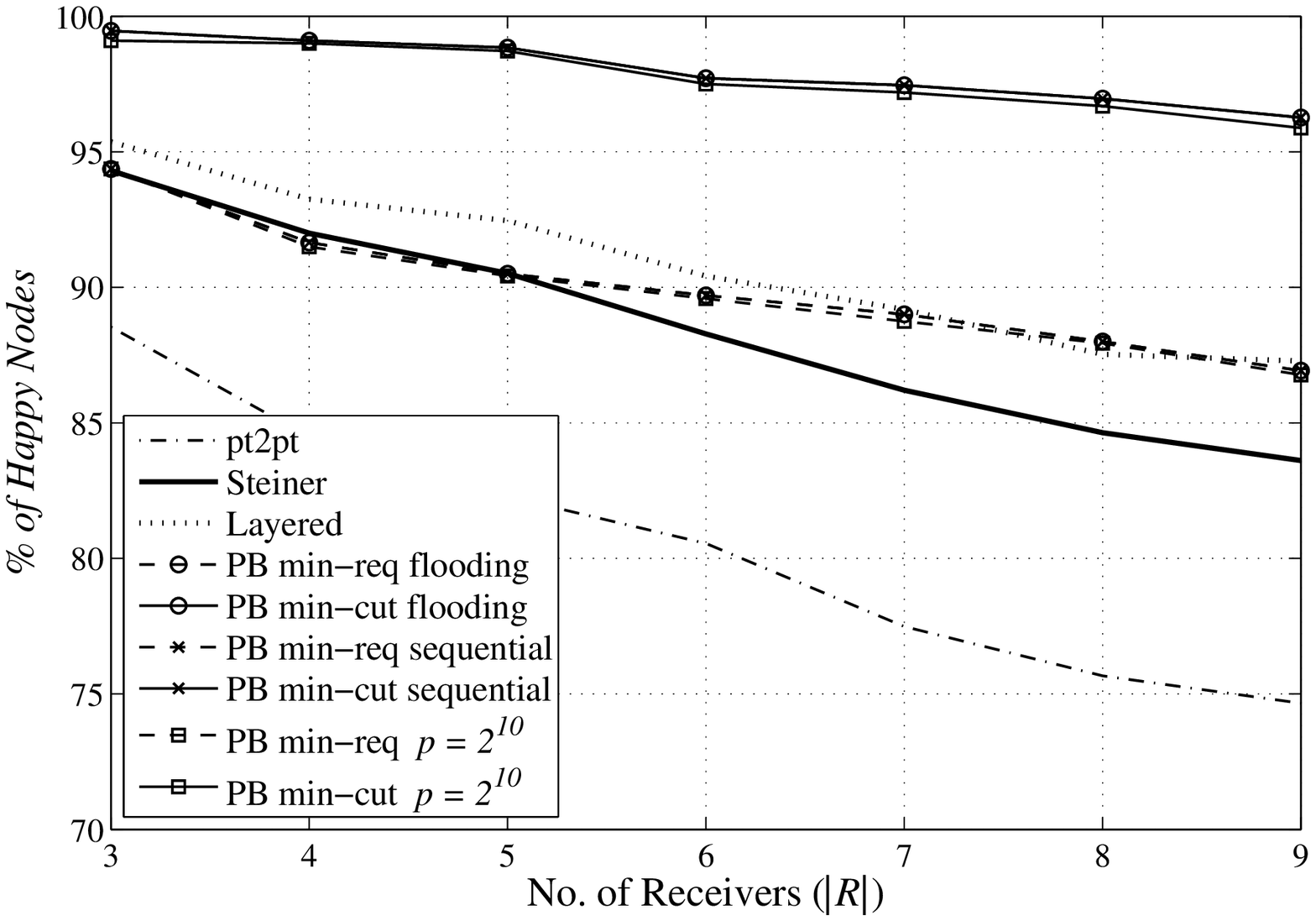}
\includegraphics[width=0.47\textwidth]{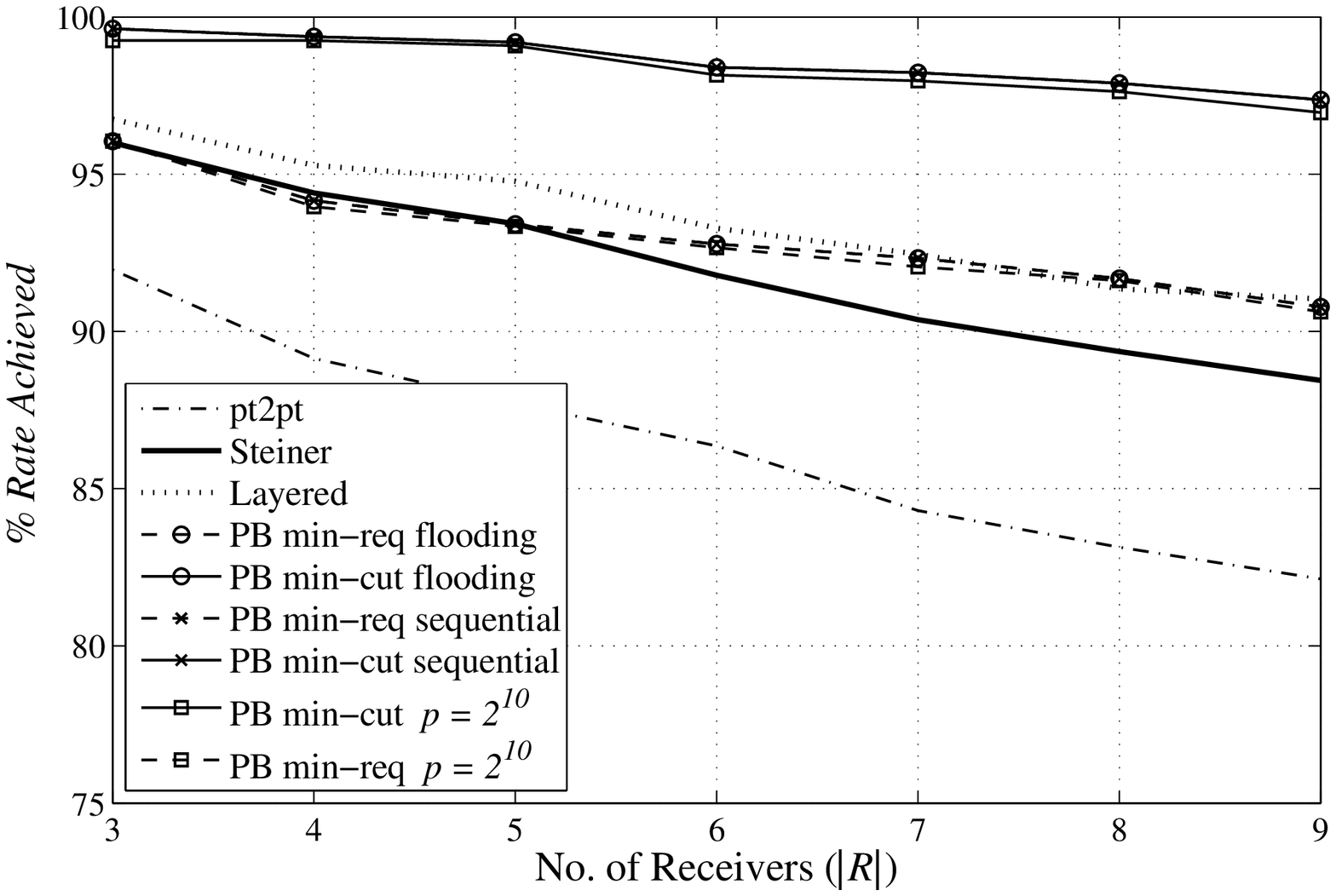}
\end{center}\vspace*{-.4cm}
\caption{Varying number of receivers in a network with 25
nodes.}\vspace*{-.3cm}\label{fig:layer2nodes25rxchanging}
\end{figure}

\begin{figure}[tbp]
\begin{center}
\includegraphics[width=0.47\textwidth]{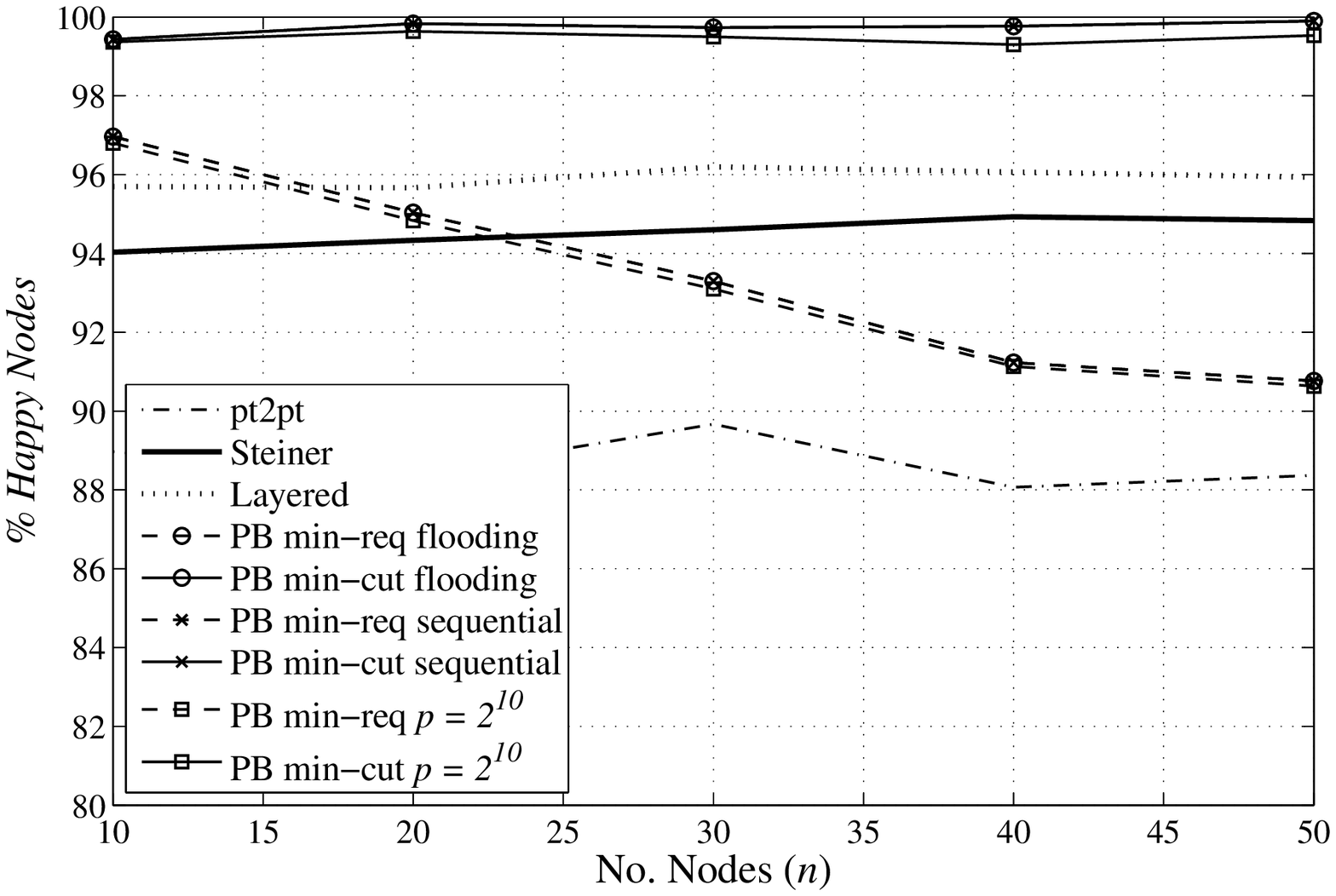}
\includegraphics[width=0.47\textwidth]{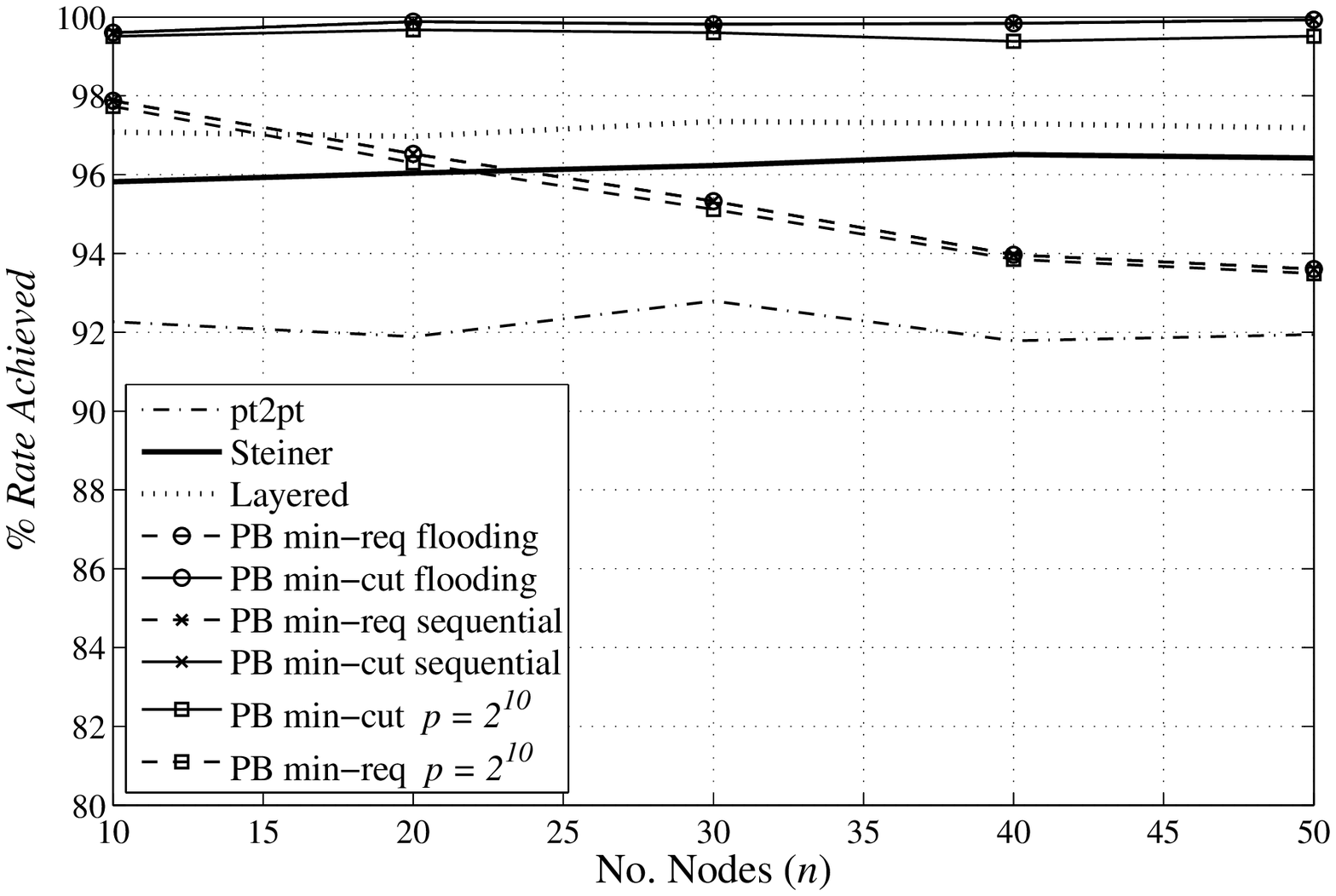}
\end{center}\vspace*{-.4cm}
\caption{Varying number of nodes in a network with 3
receivers.}\vspace*{-.3cm}\label{fig:layer2nodeschangingrx3}
\end{figure}

In Figures \ref{fig:layer2nodes25rxchanging} and \ref{fig:layer2nodeschangingrx3}, we compare the performance of the various schemes in terms of the two metrics \textit{\% Happy Nodes} and \textit{\% Rate Achieved}. We compare our pushback algorithm to that of Point-to-point Routing Algorithm (`pt2pt'), Steiner Tree Routing Algorithm (`Steiner'), and Intra-layer Network Coding Algorithm (`Layered'). We also compare the two implementations of our algorithm (flooding and sequential message passing). The flooding approaches with an infinite field size are labeled `PB min-req flooding' and `PB min-cut flooding' for min-req and min-cut criteria, respectively. The sequential message passing approaches with an infinite field size are labeled `PB min-req sequential' and `PB min-cut sequential' for min-req and min-cut criteria, respectively. Finally, we include results when a moderate field size ($p = 2^{10}$) is used. These are labeled `PB min-req $p = 2^{10}$' and `PB min-cut $p = 2^{10}$' for the min-req and min-cut criteria, respectively.

Figure \ref{fig:layer2nodes25rxchanging} shows the performance of the various schemes when we increase the number of receivers in the network. The pushback algorithm with min-cut criterion has the best performance overall. The flooding approach and the sequential message passing approach have the same performance, and furthermore, using a moderate field size of $p = 2^{10}$ yields results close to that of an infinite field size. This can be seen for both the min-cut and the min-req versions. Note that the performance of the various scheme follow a similar trend for both metrics \textit{\% Happy Nodes} and \textit{\% Rate Achieved}.



In addition, Figure~\ref{fig:layer2nodes25rxchanging} illustrates
that the gap between the min-cut version of our algorithm and
`pt2pt',  `Steiner' and `Layered' increases with the
number of receivers in the network. Note that the gap between the
min-cut and the  min-req criteria increases more slowly than the gap
between the min-cut and the other schemes.

Figure~\ref{fig:layer2nodeschangingrx3} compares the performance of
the different schemes with fixed number of receivers and varying number of nodes in the network. Note our algorithm with
the min-cut criterion outperforms the intra-layer network coding and the routing schemes. In fact, the min-cut criterion consistently achieves close to 100\% for both
\textit{\%~Happy~Nodes} and \textit{\% Rate Achieved} while the second
best scheme (`Layered') achieves at most $96$\% and $97$\% for
the two metrics.

Figure~\ref{fig:layer2nodeschangingrx3} shows that the performance
of the min-cut criterion is very robust to the
number of nodes in the network. In fact, the performance improves
as more nodes are available. However, the min-req version
degrades with the number of nodes. This is because, when using the min-req criterion, the requests from receivers with
min-cut equal to one limits the rate of other receivers. When the network
becomes large, this flooding of base layer requests has a more
significant effect on the throughput as there are more resources
wasted in delivering just the base layer. This indicates that the choice of network code can greatly impact
the overall network performance, depending on its topology and
demands. An inappropriate choice of network code can be detrimental,
shown by the min-req criterion (`PB min-req'); however, an
intelligent choice of network code can improve the performance
significantly, shown by the min-cut criterion (`PB min-cut').


\begin{figure}[tbp]
\begin{center}
\includegraphics[width=0.47\textwidth]{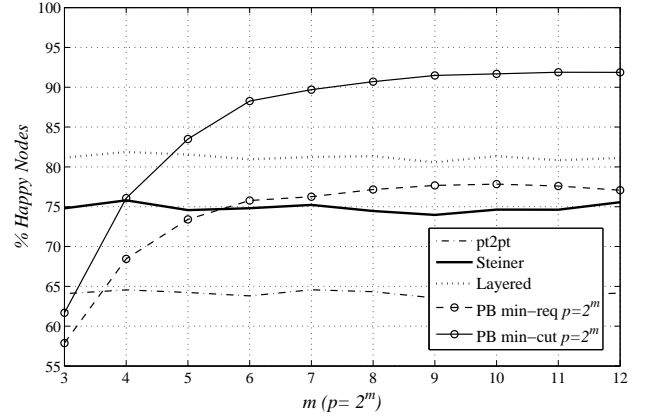}
\end{center}\vspace*{-.4cm}
\caption{Varying field size in a network with 9 receivers 25
nodes.}\vspace*{-.3cm}\label{fig:layer3nodes25rx9}
\end{figure}

\subsection{Simulation results for 3-layer multi-resolution code}
Similarly to the 2-layer case, for 3-layer multi-resolution codes,
we generated random networks to evaluate the pushback algorithm. For
each data point in the plots, we averaged 1000 trials. The min-cuts
and the in-degrees of all nodes were less than or equal to $3$. Recall
that with $3$ layers, the patterns of coding coefficients are not
sufficient for checking the decodability of incoming packets.
Instead, Gauss-Jordan elimination is necessary at every node during the
code generation stage.

Figure~\ref{fig:layer3nodes25rx9} illustrates the effect of field
size given a network of 25 nodes and 9 receivers. The pushback
algorithm with the min-cut criterion (`PB min-cut') outperforms
routing and intra-layer coding schemes (`Layered') with a field size of $p =
2^5$. Note, in terms of \textit{\% Happy Nodes}, `PB min-cut' achieves roughly $92\%$ when the
field size is large enough, while the intra-layer coding scheme
 only achieves about $82\%$.
Figure~\ref{fig:layer3nodes25rx9} also illustrates that intra-layer
coding scheme still outperforms the routing
schemes, even when optimal multicast routing is used for each layer
(`Steiner'). Our pushback algorithm achieves considerably
higher gains by performing inter-layer in addition to intra-layer
coding.
\begin{figure}[tbp]
\begin{center}
\includegraphics[width=0.47\textwidth]{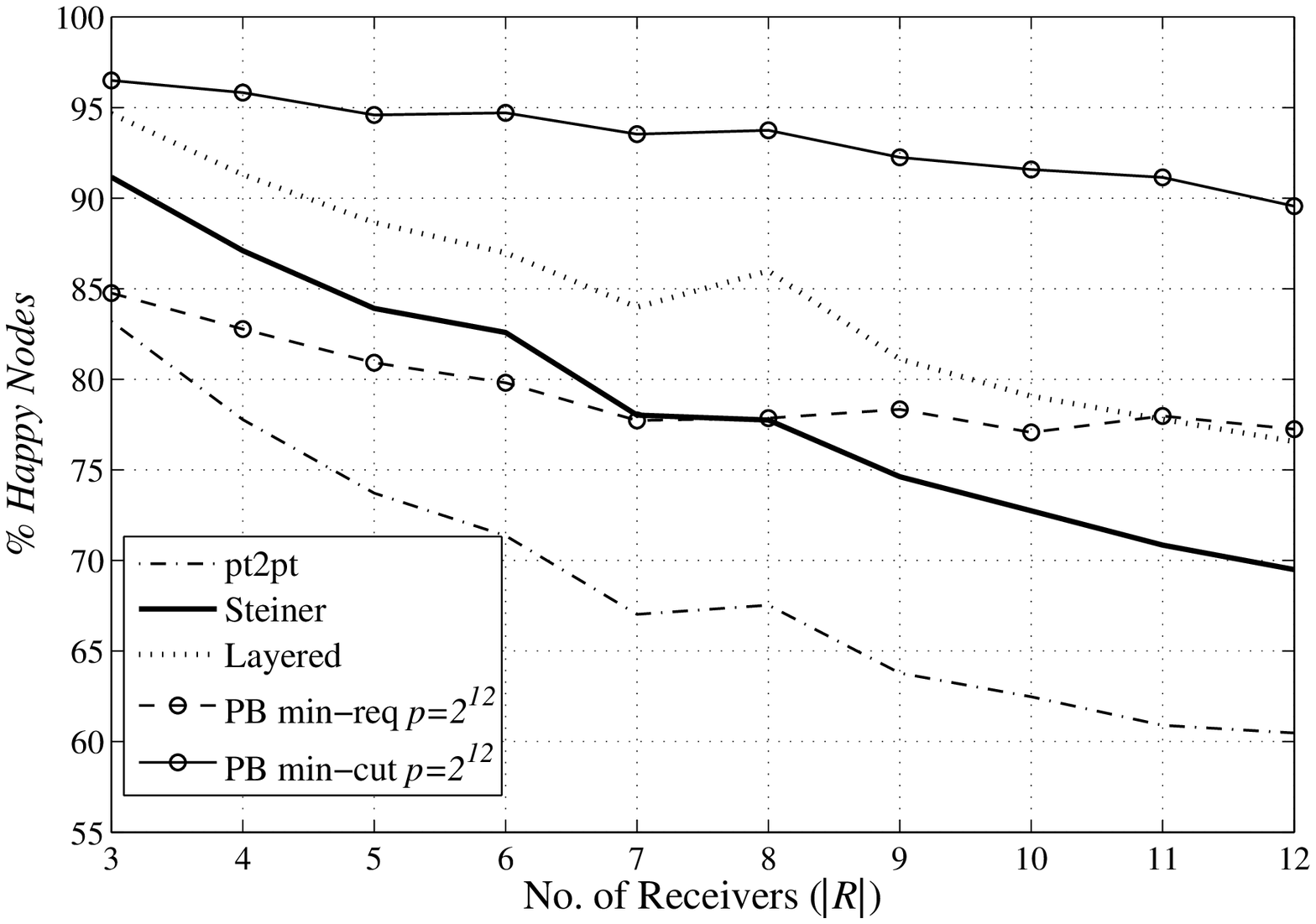}
\includegraphics[width=0.47\textwidth]{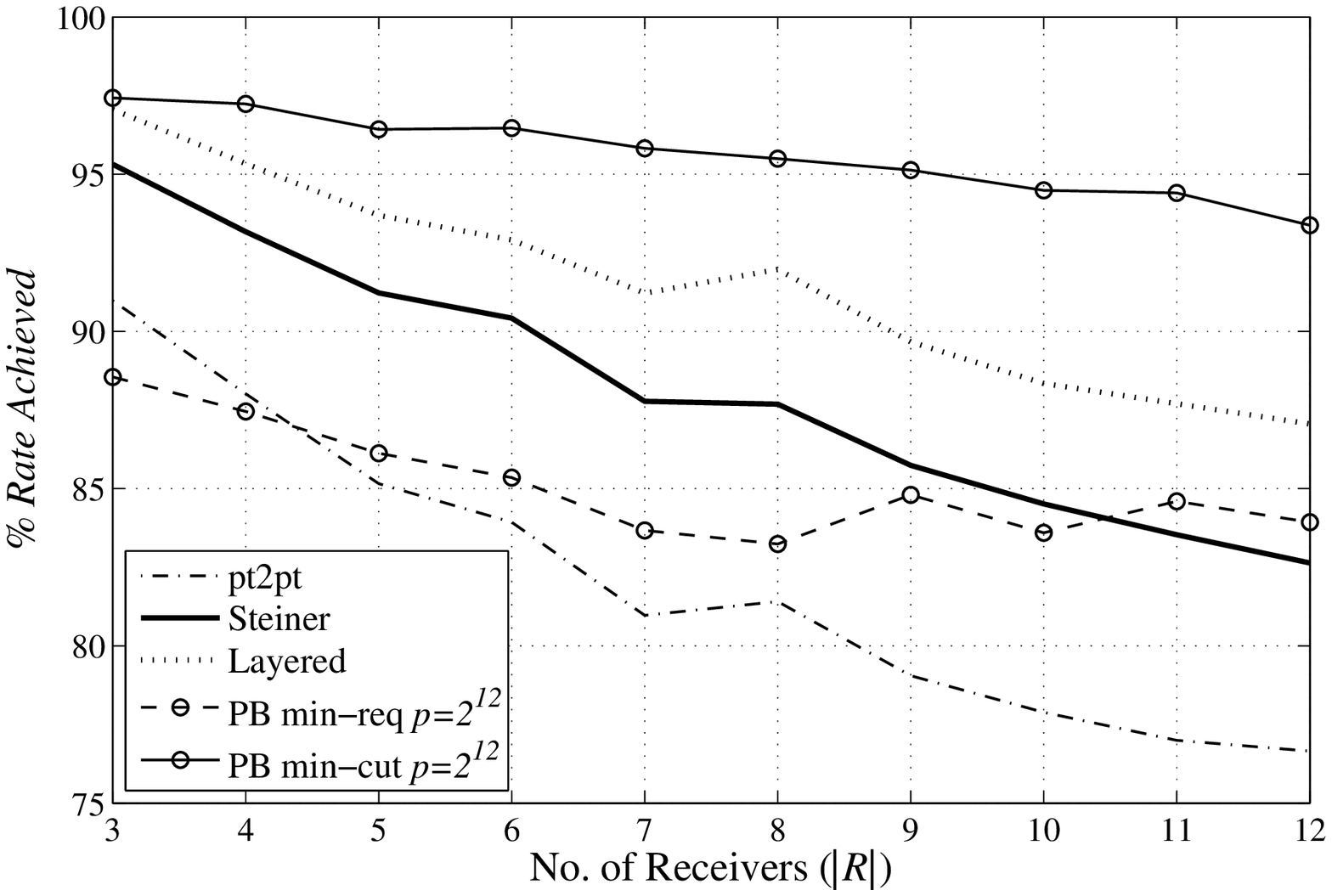}
\end{center}\vspace*{-.4cm}
\caption{Varying number of receivers in a network with 25 nodes using field
size of $2^{12}$.}\vspace*{-.3cm}\label{fig:layer3nodes25happy}
\end{figure}

As the number of receivers increases, more demands need to be
satisfied simultaneously. It is therefore expected that the overall
performance of multicast schemes will degrade with the number of
receivers. This can be observed in
Figures~\ref{fig:layer3nodes25happy}. Nonetheless, the performance
gap between our two criteria of pushback (`PB min-cut' and `PB
min-req') remains approximately constant, while the performance gain
over other schemes increases. This means that our algorithm is more
robust to changes in the number of receivers than the other schemes,
an important property for systems that aim to provide service to a
large number of heterogeneous users.

\begin{figure}[h]
\begin{center}
\includegraphics[width=0.47\textwidth]{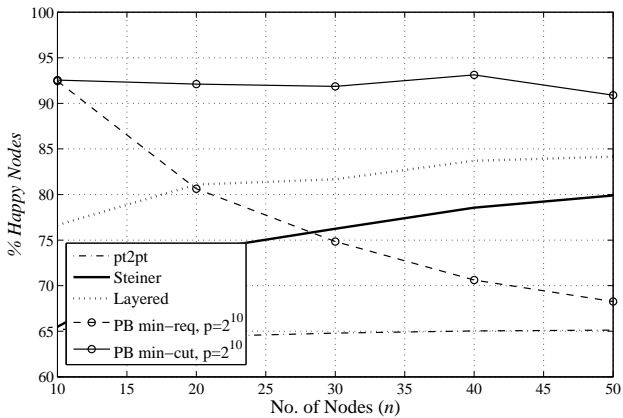}
\end{center}\vspace*{-.4cm}
\caption{Varying number of nodes in a network with 9
receivers using field size of $2^{10}$.
}\vspace*{-.3cm}\label{fig:layer3m10happy}
\end{figure}

Figure~\ref{fig:layer3m10happy} illustrates the performance of the
different schemes when we increase the number of nodes in the
network. As the number of nodes increases, there are more disjoint paths within the network for Steiner tree routing and
intra-layer coding to use. Hence the
performance of these schemes improves. The opposite behavior occurs
for the pushback algorithm with the min-req criterion, \emph{i.e.}
the \textit{\% Happy Nodes} decreases with an increase in the number
of nodes in the network. This result is similar to that of
Figure~\ref{fig:layer2nodeschangingrx3} for 2-layer case. Note that
as the number of nodes in the network increases, it becomes more
likely that a small request by one receiver suppresses higher
requests by many other receivers. Hence, pushback with the min-req
criterion quickly deteriorates in terms of \textit{\% Happy Nodes}.

\section{Conclusions and Future work}\label{sec:conclusions}
A simple, distributed message passing algorithm, called the
\emph{pushback algorithm}, has been proposed to generate network
codes for single source multicast of multi-resolution codes. With
two stages, the pushback algorithm guarantees decodability of the
base layer at all receivers. In terms of total rate achieved, this
algorithm outperforms routing schemes as well as intra-layer coding
schemes, even with small field sizes such as $2^{10}$. The
performance gain increases as the number of receivers increases and
as the network grows in size as shown by numerical simulations.

Possible future work includes the addition of a third
\emph{complaint} stage, in which receivers whose requests have not
been satisfied pass another set of requests to their parents,
signaling their desire for more. In generating new codes, parent
nodes must take into account the new updated
requests, while maintaining decodability at receivers which
did not participate in the complaint stage. It is important to determine what the complaint messages should be, and to assess the improvements that can be achieved with such an additional stage.

Another important extension is to apply this algorithm in
wireless/dynamic multicast settings. The \emph{flooding} approach (Section \ref{sec:simulations}) is applicable to such
settings, as changes in the network can be handled by new messages
to the neighboring nodes. An important extension is to study the performance and the convergence of this flooding approach in dynamic settings.

Lastly, in the pushback algorithm, \emph{rate} is the message sent
by nodes to their parens, \emph{i.e.} each node signals how many
layers down-stream receivers can or want to receive. It may be
possible to extend the message to include other constraints, such as
power (decoding power), delay, and reliability.


\bibliographystyle{IEEEtran}
\bibliography{References}
\end{document}